\documentstyle[11pt,newpasp,twoside,epsf,psfig]{article} 
\def\edcomment#1{\iffalse\marginpar{\raggedright\sl#1\/}\else\relax\fi} 
\reversemarginpar

\begin{document} 
\def\gta{ \lower .75ex \hbox{$\sim$} \llap{\raise .27ex \hbox{$>$}} }
\def\lta{ \lower .75ex\hbox{$\sim$} \llap{\raise .27ex \hbox{$<$}} }
 
\title{Testing Comptonization models using\\
 {\it Beppo}SAX observations of Seyfert 1 galaxies}  
\author{P.O. Petrucci}
\affil{$^1$Osservatorio Astronomico di Brera, Milano, Italy}
\author{F. Haardt$^{2}$, L. Maraschi$^{1}$, P. Grandi$^{3}$,
 J. Malzac$^{1}$, G. Matt$^{6}$, F. Nicastro$^{3,4,5}$, L. Piro$^{3}$,
 G.C. Perola$^{6}$, A. De Rosa$^{3}$.}

\affil{$^1$Osservatorio Astronomico di Brera, Milano, Italy;
$^2$Universit\'a dell'Insubria, Como, Italy; 
$^3$IAS/CNR, Roma, Italy; 
$^4$CfA, Cambridge Ma., USA;
$^5$Osservatorio Astronomico di Roma, Roma, Italy;
$^6$Universit\'a degli Studi ``Roma 3'', Roma, Italy}

\begin{abstract} 
We used high quality BeppoSAX data of 6 Seyfert galaxies to test
realistic thermal Comptonization models. Our main effort was to adopt
a Comptonization model taking into account the anisotropy of the soft
photon field. The best fit parameter values of the temperature and
optical depth of the corona and of the reflection normalization
obtained fitting this class of models to the data are substantially
different from those derived fitting the same data with the power law
+ cut--off model commonly used. The two models also provide different
trends and correlation between the physical parameters, which has
major consequences for the physical interpretation of the data
\end{abstract}

\section{Introduction} 
The broad band X-ray spectra (2-300 keV) of Seyfert 1 galaxies are
generally well fitted by a cut-off power law continuum + reflection
This model depends on three parameters: the spectral index $\Gamma$,
the high energy cut-off $E_c$ and the reflection normalization
$R$. The physical interpretation of these parameters is generally done
in the framework of thermal Comptonization mechanism, which is
commonly believed to be the origin of the X-ray emission of Seyfert
galaxies. Using approximate relations (cf. section 2), it is then
possible to derive, from values of $\Gamma$ and $E_c$, values of the
temperature $kT_e$ and optical depth $\tau$ of the ``comptonizing''
hot plasma (the so-called corona).\\ If such approximations are
sufficient in isotropic geometries, where Comptonization process
produces roughly cut--off power law spectra, strong discrepancies may
appear in anisotropic ones, especially for small optical depth and
large temperature. The contribution of the first scattering order to
the outgoing flux is then significantly reduced producing a spectral
break (the so-called anisotropy break) in the spectrum.  Consequently
the physical parameter obtained fitting this class of models to broad
band X-ray spectra are substantially different from those derived
fitting the same data with the power law + cut--off model. In a recent
paper (Petrucci et al., P00), we have applied accurate Comptonization
models to fit high quality data of the BeppoSAX long look at NGC~5548
and we have underlined the importance of the anisotropic effects.  In
the same spirit of P00, we extend our application of accurate
Comptonization models to a larger sample of objects. We report in the
following the results of this study.

\section{Data and model fitting}
\label{result}
From the complete sample of Seyfert 1 observed by BeppoSAX we have
selected six objects (NGC 5548, IC 4329A, NGC 4151, ESO 141-G55, Mkn
509 and NGC 3783) whose observations have high signal-to-noise ratios
and hard spectra yielding a good detection in the PDS instrument.
Concerning NGC 4151, we have used the observations of July 1996 and
January 1999.\\ We have fitted the data using two different models for
the primary continuum: (1) an exponentially cut--off power law plus a
reflection component from neutral material (PEXRAV model of XSPEC,
Magdziarz \& Zdziardski, 1995) and (2) a thermal Comptonization
spectrum from a disk+corona configuration in slab geometry (code of
Haardt (1994, hereafter H94). The fit parameters of H94 are the
temperature of the corona $kT_e$, its optical depth $\tau$, the
temperature of the disk $kT_{bb}$ (assuming a black body soft
emission) and the reflection normalization $R$. On the other hand, the
PEXRAV continuum depends only on 3 parameters: the e--folding energy
of the cut--off power law $E_{\rm c}$, the photon index $\Gamma$ and
the reflection normalization $R$. The temperature $kT_e$ inferred from
the PEXRAV fits is simply computed as $kT_e\equiv E_{\rm c}/2$,
keeping in mind that such approximation roughly holds for $\tau\lta
1$. For $\tau \gg 1$, $kT_e\equiv E_{\rm c}/3$ would be more
appropriate. Knowing the temperature, the spectral index derived from
the PEXRAV fit can be used to determine the optical depth using the
following relation $\displaystyle\Gamma-1
\simeq\left[\frac{9}{4}+\frac{m_ec^2}{kT_e \tau (1+\tau
/3)}\right]^{1/2}-\frac{3}{2}$ (Shapiro, Lightman \& Eardley,
1976). This equation is valid for $\tau> 1$, and we have checked {\it
a posteriori} that such condition is roughly matched in all cases.\\

\section{Results}
\label{result}
\subsection{Comparison between the slab model and PEXRAV}
For all objects, the estimated corona temperatures (respectively
optical depths) from PEXRAV are substantially smaller (respectively
larger) than those inferred with an anisotropic Comptonization model
(cf. Fig. 1a). Large differences (up to a factor 8) are
found between the two temperature estimates.\\
The reflection normalizations obtained with H94 cluster around 1,
except for two extreme states of NGC 4151.  They are, in all cases but
one (ES0 141-G55), larger than those found using the simple cut--off
power-law model (cf. Fig. 1b). In some cases we obtain
differences of factors 4--5. \\
\begin{figure}[h]
\plottwo{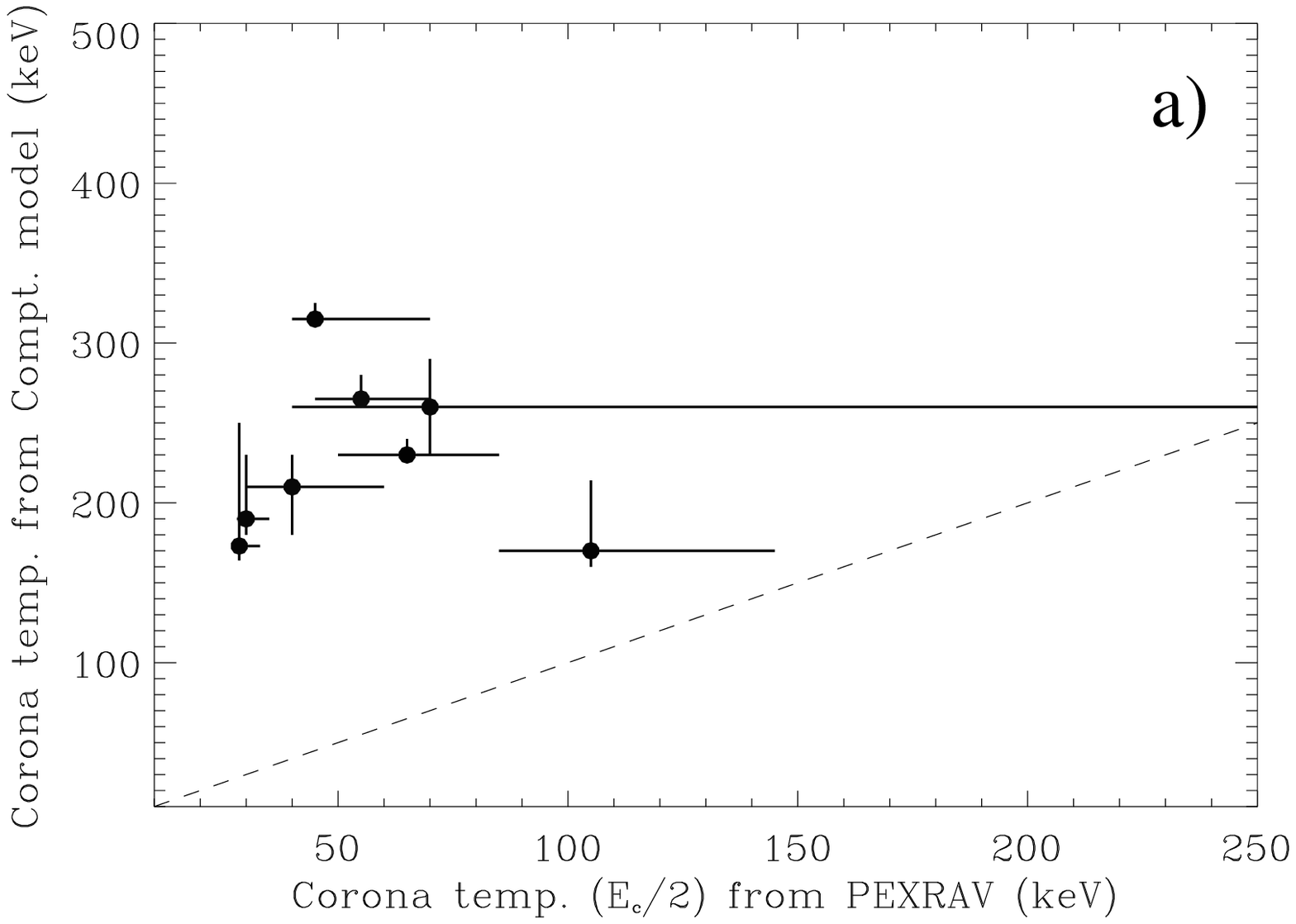}{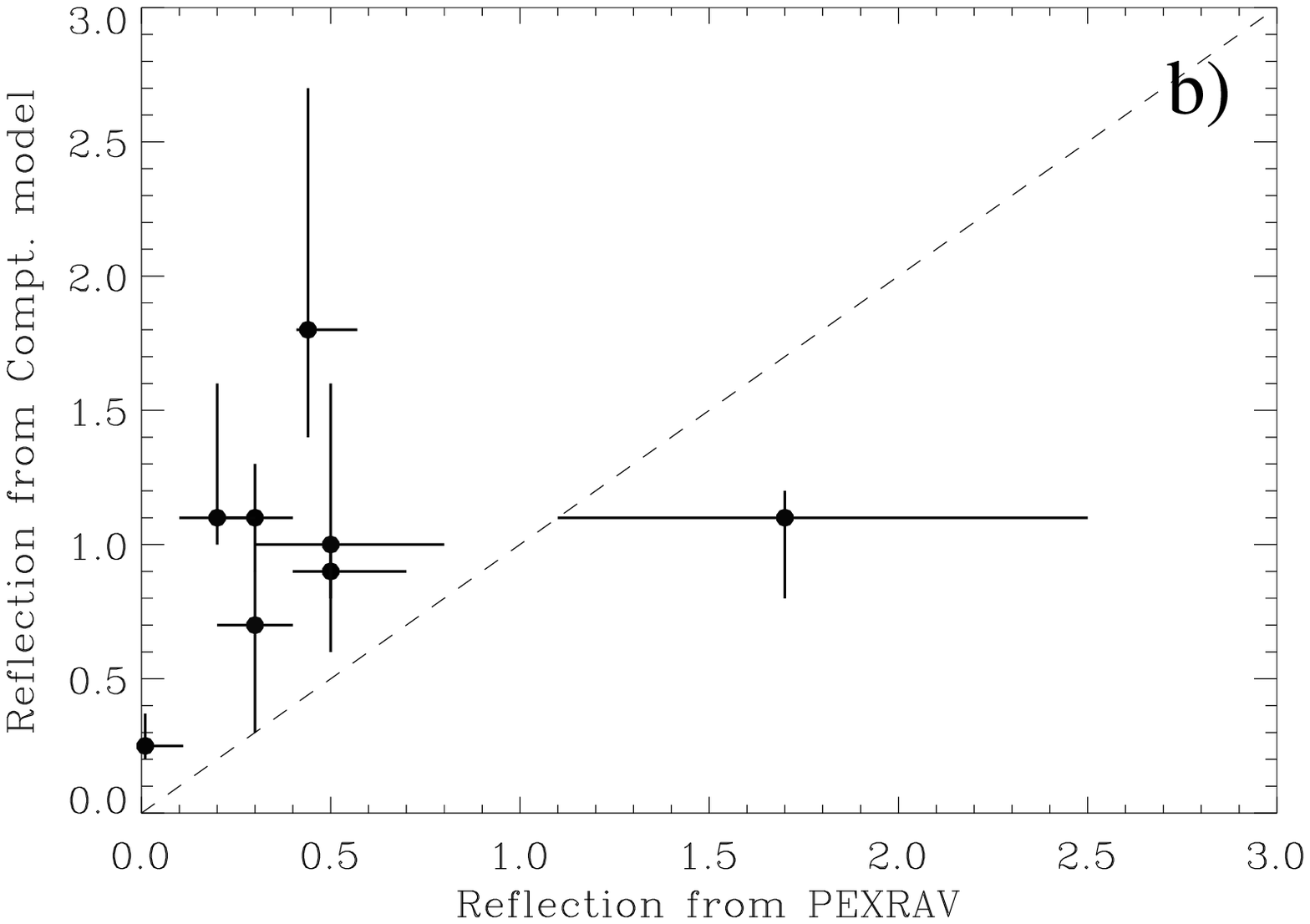}
\caption{Slab corona temperature and reflection normalization versus
PEXRAV estimates}
\end{figure}
The reason of these differences is relatively simple. Within PEXRAV
type models, the slope of the power law, determined with small errors
by the LECS and MECS data, cannot change, by hypothesis, at higher
energies. A cut--off around 100 keV is then required to fit the PDS
data. In Comptonization models, the LECS and MECS data determine the
slope below the anisotropy break. Above this break the intrinsic
spectrum is steeper. It can thus fit the PDS data without an
additional steepening beyond 100 keV, allowing for a larger
temperature (and consequently a smaller optical depth to keep,
roughly, the same power law slope) and a larger value of the
reflection component normalization. \\

\section{Discussion}
\label{discu}
In Fig. 2a and 2b the values of $\tau$ vs. $kT_e$, obtained for the 6
sources using the spectral models H94 and PEXRAV, are compared with
the theoretical relations expected for a plane and hemispherical
Comptonizing region in energy balance (the temperature and optical
depth then satisfy a definite and univocal relation which corresponds
to roughly constant Compton parameters $\displaystyle y \simeq
4\left(\frac{kT_e}{m_ec^2}\right)\
\left[1+4\left(\frac{kT_e}{m_ec^2}\right)\right] \tau (1+\tau )$.)\\
\begin{figure}[h]
\plotone{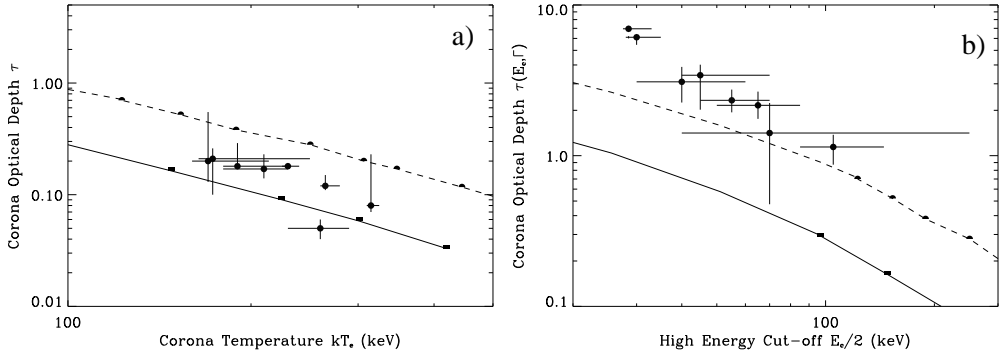}
\caption{Optical depth $\tau$ versus temperature $kT_e$ from (a) H94
and (b) PEXRAV fits.  The theoretical relations between $\tau$ and
$kT_e$ for a plane and hemispherical Comptonizing region in energy
balance, in solid and dashed line respectively (from Stern et al.,
1995), are shown for comparison.}
\end{figure}
The best fit results obtained with H94 (Fig. 2a) are relatively close
to the theoretical expectations for the slab case even if they tend to
fall preferentially above the solid line. The data therefore indicate
a Comptonization parameter larger than for a pure slab geometry, that
is a more ``photon-starved'' configuration.
The values of $kT_e$ and $\tau$ obtained with PEXRAV (cf. Fig. 2b)
show the same trend as found with H94 that is larger values of $\tau$
for smaller values of $kT_e$. They are also above the theoretical
expectation for a slab and even above the theoretical expectation for
a hemisphere. Therefore this set of parameters suggests, for each
source of our sample, a configuration more ``photon-starved'' than a
hemisphere.\\ An important difference with respect to the slab model
is that the optical depths derived from PEXRAV fits are generally
larger than 1. The corona should then be optically thick reducing or
canceling the effects of anisotropy. Therefore also the PEXRAV model
has an internal consistency. However a corona with large optical depth
may wash out discrete features from the underlying disk (e.g. lines
and reflection itself) more than would be desirable. This problem
could be alleviated if the corona was "patchy". \\ We conclude that
both a hot, optically thin corona with significant anisotropic effects
and a less hot, optically thick, patchy corona with negligible
anisotropy are consistent with the available data.

\subsection{Correlations between physical parameters}
A correlation between the reflection normalization $R$ and the photon
index $\Gamma$ has been claimed by Zdziarski et al. (1999, hereafter
Z99) from the study of a large number of GINGA observations of Seyfert
and galactic black hole objects. Yet, the validity of this correlation
is still under debate since $R$ and $\Gamma$ are strongly correlated
in the fitting procedure (cf. for example Vaughan \& Edelson, 2000).\\
We have plotted in Fig. 3a the reflection normalization $R$ versus the
photon index $\Gamma$ obtained with PEXRAV.  Our data don't show a
clear correlation between the two parameters. However our sample is
biased in favor of objects with hard spectra, to ensure a good
detection in the PDS instrument. When the 13 objects actually observed
by BeppoSAX are taken into account a stronger correlation is observed
(Matt, 2000), in agreement with Z99. In the case of the H94 model we
have no simple way of characterizing the spectral shape. Also results
are available for only 6 sources and we cannot include the other
Seyfert 1s observed by BeppoSAX.  Therefore a direct comparison with
the Z99 correlation is not possible.\\
\begin{figure}[h]
\plotone{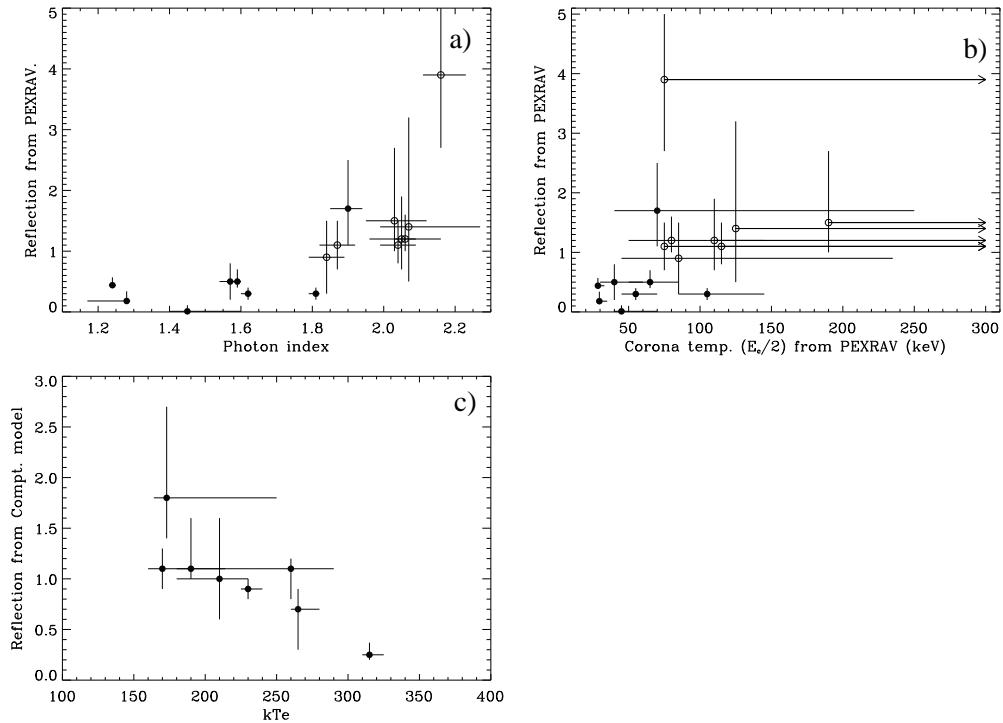}
\caption{a) Reflection $R_{PEXRAV}$ versus PEXRAV photon index
$\Gamma$. b) Reflection component $R_{PEXRAV}$ versus PEXRAV corona
temperature $E_c/2$ from PEXRAV. The open circles are from Matt (2000)
while the solid ones are the results of this work. c) $R_{slab}$ versus
slab corona temperature $kT_e$}
\label{Rvsgamma}
\end{figure}
Z99 interpret the correlation between $R$ and $\Gamma$ in the
framework of thermal reprocessing models where $R$ is directly
proportional to the solid angle subtended by the cold matter
surrounding the corona. In this case, the larger $R$ and the larger
the cooling of the corona resulting in softer spectra (larger
$\Gamma$).\\ It is interesting that the BeppoSAX data can provide the
temperatures associated with sources with different values of $R$ or
$\Gamma$, which were not available from the GINGA data used by
Z99. The result is shown in Fig. 3b where we have plotted $R_{PEXRAV}$
vs.  the corona temperature $E_C/2$. This plot suggests a {\it
positive} correlation between the two parameters, that is, the
temperature of the corona is {\it larger} for larger values of the
reflection normalization.  A plot of $E_C/2$ vs. $\Gamma$ shows that
steeper spectra correspond to higher temperatures and (as a corollary)
to lower optical depths.\\ We note that the slab model analysis
suggests different trends.  If we plot $R$ vs. $kT_e$ for the latter
model we find that sources with larger $R$ tend to have lower $kT_e$
(cf. Fig. 3c). Looking also at Fig 2 it appears that this model yields
a smaller range of optical depths for the sources in the sample, thus
sources with flatter spectra have larger temperatures, a behavior
opposite to what is found with PEXRAV.\\ It is interesting to note
that, in the interpretation of Z99, the $R_{PEXRAV}$--$E_c/2$
correlation of Fig. 3b suggests that the corona temperature is larger
for larger cooling. Such behavior is expected in pair dominated
thermal plasma in pair equilibrium (i.e. where pair creation is
balanced by pair annihilation). Indeed, in this case, an increase of
the cooling corresponds to a decrease of the number of particles in
the hard tail of the thermal particle distribution, and thus to a
decrease of the pair production rate. The number of particles in the
thermal bath thus decreases, i.e. $\tau$ decreases. If the heating is
kept constant, the available energy is now shared among less
particles, resulting in an increase of the temperature.\\ On the
contrary, the anticorrelation between $R$ and $kT_e$ obtained with the
slab corona model (Fig. 3c) is naturally expected in the framework of
reprocessing models but for low pair density corona, in complete
opposition with the conclusions deduce from our PEXRAV fits.\\

\section{Conclusion}
The aim of this paper was to test thermal Comptonization models over
the high signal to noise BeppoSAX observations of a sample of six
Seyfert 1. We use two types of model: a detailed Comptonization code
in slab geometry, which treats carefully the anisotropy effects in
Compton processes, and a simple cut--off power law plus reflection
model (PEXRAV model of XSPEC). The latter is a relatively good
approximation to Comptonization spectra in isotropic geometry and/or
optically thick corona. The main results of this work are the
following:\\ \indent{$\bullet$ The} data are well fitted by both
models and there is no statistical evidence for a model to be better
than the other. Both models give results in agreement with a X-ray
source geometry more ``photon starved'' than the slab case.\\
\indent{$\bullet$ There} are strong differences between the best fit values
of the temperature and optical depth of the corona and the reflection
normalization obtained with the two models. We generally obtained
larger corona temperature, smaller optical depth and larger reflection
normalization with the slab geometry in comparison to $PEXRAV$\\
\indent{$\bullet$ The} two models leads to strong different relationships
between physical parameters. For instance, we obtain a correlation
between the reflection normalization $R$ and the corona temperature
with PEXRAV, and an anticorrelation with the slab corona model. This
has major consequences for the physical interpretation of the data.\\
Forthcoming observations with CHANDRA, XMM-Newton and INTEGRAL are
expected to bring substantial progress to discriminate between these
two types of models for the high energy continuum of Seyfert
galaxies.\\

\noindent
{\sl Acknowledgements:} POP acknowledges a grant of the European
Commission under contract number ERBFMRX-CT98-0195 (TMR network
"Accretion onto black holes, compact stars and protostars").


\begin{references}
\small
\reference Haardt, F. 1994, PhD dissertation, SISSA, Trieste 
\reference Haardt, F. \& Maraschi, L.  1991, \apjl, 380, L51 
\reference Magdziarz, P. \& Zdziarski, A. 1995, MNRAS, 273, 837
\reference Matt, G., "X-Ray Astronomy '99" proceedings, 1999, astro-ph/0007105
\reference Petrucci, P. et al. 2000, \apj, 540, 131
\reference Shapiro, S.  et al. 1976, ApJ, 204, 187
\reference Stern, B. et al. 1995, \apjl, 449, L13
\reference Zdziarski, A., et al. 1999, MNRAS, 303, L11
\reference Vaughan, S. \& Edelson, R., ApJ, in press, astro-ph/0010274 
\end{references}
\end{document}